# Electrode Polarization in Glassy Electrolytes: Large Interfacial Capacitance Values and Indication for Pseudocapacitive Charge Storage

C. R. Mariappan, T.P. Heins, B. Roling

Department of Chemistry, University of Marburg, Hans-Meerwein-Strasse, D-35032 Marburg, Germany.

*Abstract*

*We study the electrode polarization behaviour of different Na-Ca-phosphosilicate glasses by measuring the differential capacitance between blocking Pt electrodes. At low applied dc bias voltages, we detect a linear capacitance regime with interfacial capacitance values considerably larger than expected from mean-field double layer theories and also considerably larger than found for ionic liquids with similar ion concentrations. With increasing bias voltages, the differential interfacial capacitance exhibits a maximum around 1 V and a strong drop at higher voltages. We suggest that these features are caused by pseudocapacitive processes, namely by the adsorption of mobile $Na^+$ ions at the electrodes. While pseudocapacitive processes are well known in liquid electrochemistry, more detailed studies on solid electrolytes should offer perspectives for improved energy storage in solid-state supercapacitors.*

## 1. Introduction

Ion transport in solid materials, such as defective crystals, glasses, polymers and nanocomposites, plays an important role for various types of solid-state electrochemical cells, such as thin-film batteries, supercapacitors, fuel cells and electrochemical sensors. The experimental characterisation of ion transport processes is often done by taking ac impedance spectra of a solid sample between blocking metal electrodes, e.g. platinum or gold electrodes. In this case, the high-frequency part of the impedance spectra can usually be attributed to bulk ion transport in the solid material, while the low-frequency part reflects blocking of the mobile ions by the metal electrode. This low-frequency phenomenon is called 'electrode polarization' [1, 2]. The blocking leads to an accumulation or depletion of mobile ions close to the electrodes, which is reflected in a strong increase of the frequency-dependent real part of the measured capacitance with decreasing frequency [1, 2].

The motivation for studying electrode polarisation phenomena is twofold: (i) The electrode polarization part of the impedance spectra should contain additional information about the ion transport properties of the solid electrolyte, e.g. information about the number density of mobile ions [3]. (ii) The formation of ion accumulation/depletion layers may be used to store electrical energy in supercapacitors [4, 5]. However, progress in basic science as well as in supercapacitor applications has been impeded by the current poor theoretical understanding of electrode polarisation in disordered solid electrolytes.

Pioneering theoretical work in this field was done in the 1950s and 1960s by Macdonald, Friauf, Ilschner, Beaumont, and others [6-8]. They used mean-field approaches analogous to the classical Gouy-Chapman theory for diluted electrolyte solutions. These approaches are based on the assumption that a mobile ion interacts with an average field generated by the electrode and by other mobile ions. While this assumption seems reasonable for solid electrolytes with low number densities of point defects, the applicability to disordered solids with high mobile ion concentrations is far from obvious. In this type of solids, local interactions between mobile ions should play an important role for the structure of the interfacial layers.

Nevertheless, a large number of frequency-dependent capacitance data of ion-conducting glasses and polymers were traditionally analyzed and interpreted utilizing the above mean-field approaches. From the capacitance spectra, values for the Debye length were extracted, and these values were used for calculating the number density of mobile ions. Often, the obtained number densities were considerably lower than the total ion content of the sample [3, 9, 10].

However, recent theoretical and experimental work on the double layer capacitance of ionic liquids (ILs), another class of concentrated electrolytes, provides strong evidence that this kind of interpretation of capacitance data is not appropriate. The differential capacitance per unit area, $C_A$, is defined as:

$$C_A = \frac{d\sigma}{d\Delta\varphi} \quad (1)$$

where $\sigma$ is the electrode charge per unit area and $\Delta\varphi$ the electrical potential difference between electrode and bulk of the electrolyte. Molecular dynamic simulations of charged spheres between metal electrodes reveal that the linear differential capacitance obtained for small potential differences $\Delta\varphi$ can be considerably smaller than the Debye capacitance $C_D$ [11, 12]:

$$C_D = \frac{\varepsilon_0 \varepsilon}{L_D} \quad (2)$$

Here, $\varepsilon_0$ and $\varepsilon$ denote the vacuum permittivity and the relative permittivity of the IL, respectively. The Debye length $L_D$ is given by:

$$L_D = \sqrt{\frac{\varepsilon_0 \varepsilon\, k_B T}{(N_{V,+} + N_{V,-})e^2}} \quad (3)$$

with $k_B$ and $T$ denoting Boltzmann's constant and the temperature, while $N_{V,+}$ and $N_{V,-}$ are the bulk number densities of cations and anions, respectively. Typical Debye lengths for ionic liquids are about $0.5\ \overset{\circ}{A}$, i.e. smaller than the radii of cations and anions. This indicates that the ionic radii play an important role for the structure and capacitance of double layer.

Typical experimental capacitance values for ionic liquids close to the point of zero charge are in the range from 10-30 μF/cm² [13-14]. At higher electrode charges, the differential capacitance depends clearly on $\Delta\varphi$ [13-15]. The physical origin of the potential dependence has been discussed by different authors, in particular by Kornyshev and coworkers, in terms of overscreening and lattice saturation effects [16-19].

To our knowledge, detailed experimental studies of the potential-dependent interfacial capacitance of solid electrolytes between blocking electrodes have not been published so far. In this Letter, we present the results of such studies on ion conducting glasses, namely on sodium calcium phosphosilicate glasses of different compositions. These glasses exhibit bioactive properties, and their bioactivity can influenced by electrical polarization between blocking electrodes [20, 21]. Our results demonstrate that the differential capacitance of these glasses exhibits three remarkable features: (i) After preparing a fresh glass sample with freshly sputtered Pt electrodes and after heating the sample to the desired temperature, the differential capacitance decreases with time before reaching a constant value. (ii) At low applied voltages, the differential capacitance is about one order of magnitude higher than found for ILs with similar ion densities (iii) At dc bias voltages around 1 V, the differential capacitance exhibits a maximum and drops strongly at higher voltages.

**2. Experimental**

Na-Ca-phosphosilicate glasses of compositions (mole%) 46.4 $SiO_2$ – 25.2 $Na_2O$ – 25.2 CaO – 3.2 $P_2O_5$ (46S4), 42.1 $SiO_2$ – 26.3 $Na_2O$ – 29.0 CaO – 2.6 $P_2O_5$ (42S6) and 44.6 $SiO_2$ – 31.2 $Na_2O$ – 23.2 CaO – 1.0 $P_2O_5$ (45S11) were prepared by heating dry mixtures of $Na_2CO_3$, $CaCO_3$, $SiO_2$, and $(NH_4)_2HPO_4$ in a Pt crucible and melting them in an electric furnace between 1623 K and 1673 K for 2 hours. After complete homogenization, the melts were poured into preheated stainless steel molds with a cylindrical shape (diameter about 20 mm). The obtained bulk glass samples were then annealed 40 K below their respective glass transition temperatures (determined by differential scanning calorimetry) for 10 hours. The annealed glass discs were cut into slices with a thickness of about 700 – 800 μm using a high-precision cutting machine (Struers Accutom–5) equipped with a diamond saw blade. Finally, both faces of the glass slices were polished with SiC (mesh 1200) by high-precision grinding using a lapping machine (Logitec PM5). For the ac impedance measurements, Pt electrodes were sputtered onto both faces of the samples.

The ac impedance spectra of the glasses were measured in a frequency range from 0.01 Hz to 1 MHz using a Novocontrol Alpha-AK impedance analyzer equipped with a POT/GAL 15V/10A electrochemical interface. A low rms ac voltage of 10 mV and variable dc bias voltages up to 14 V were applied to the samples. The sample temperature was controlled (±0.02 K) by the Novocontrol Quatro Cryosystem.

**3. Results and Discussion**

For observing the electrode polarization in the frequency range of our measurements (ν > 0.01 Hz), the samples had to be heated to elevated temperatures. Fig. 1 depicts the real part of frequency-dependent ac capacitance C'(ν) (per unit area) of the 46S4 glass measured at T = 573 K without applying a dc bias voltage. The capacitance was calculated by using the following expression:

$$C'(\nu) = \frac{1}{2\pi \cdot \nu} \cdot \frac{Z''(\nu)}{[Z'(\nu)]^2 + [Z''(\nu)]^2} \cdot \frac{1}{A} \quad (4)$$

where A is the area of the sample faces, and Z'(ν) and Z''(ν) are the real and imaginary part of the complex impedance, respectively. At high frequencies above 10 kHz, a capacitance plateau is found reflecting the bulk capacitance of the sample. With decreasing frequency, C'(ν) increases strongly, and below 0.1 Hz, there is a levelling-off into a plateau regime reflecting the interfacial capacitance.

When we carried out measurements on a freshly prepared sample with freshly sputtered Pt electrodes, we observed a drop of the interfacial capacitance with increasing measurement time. This is shown in Fig. 2 for the 46S4 glass at T = 573 K and ν = 0.01 Hz. The capacitance C'(0.01 Hz) decreases exponentially with time and reaches a stable value after about 3 hours.

Therefore, measurements with applied dc bias voltage were not started before stable ac capacitance values were detected.

In Fig. 3, the frequency-dependent capacitance of the 42S6 glass is shown for various applied dc bias voltages, $V_{dc}$. In the bulk capacitance regime, the capacitance is independent of $V_{dc}$, but in the frequency range below 10 kHz, the dc bias voltage exerts a significant influence on the capacitance. At dc bias voltages below 3 V, reliable capacitance data could be obtained down to $10^{-2}$ Hz. At higher dc bias voltages, the capacitance relaxation shifted to higher frequencies, and the low-frequency part of the spectra became noisy. The open circles in Fig. 3 denote the lowest frequencies where noise-free data could be obtained. The capacitance values at these frequencies were used for plotting the voltage dependence of the differential interfacial capacitance in Fig. 4 (closed symbols).

Although the exact capacitance values depend on the glass composition, we observe several common features (i) At dc bias voltages below 100 mV, there is a linear regime with a virtually bias voltage-independent interfacial capacitance. Here, we obtain large capacitance values in a range from 100 to 300 μF/cm$^2$. (ii) Above 100 mV, the differential interfacial capacitance increases with increasing dc bias and reaches a maximum around 1 V. (iii) Above 1 V, the differential interfacial capacitance decreases strongly with increasing dc bias.

In order to test whether these voltage-dependent changes in the differential capacitance are reversible or irreversible, we performed the following test: After taking a capacitance spectrum at a specific bias voltage, a second spectrum without applied dc bias was recorded. Then, the next spectrum was taken with a higher dc bias, followed again by a measurement without dc bias. The open symbols in Fig. 4 show the differential interfacial capacitances obtained under zero dc bias, directly after having performed a measurement at $V_{dc}$ (shown as closed symbol). If the changes in the differential interfacial capacitance were completely reversible, then the open symbols for a specific glass should be all at the same capacitance value. As seen from the figures, the irreversible effects due to the dc bias are weak in the case of the 46S4 and the 42S6 glass, while they are more pronounced in the case of the 45S12 glass. Nevertheless for all three glasses, the major part of the dc bias voltage dependence of interfacial capacitance is due to reversible effects.

In the following, we discuss the implications of these experimental findings. First of all, we note that the linear capacitance values of our glasses obtained at low voltages are by about one order to magnitude higher than those found in ionic liquids with similar ion densities and Debye lengths. What values for the interfacial capacitance do we expect when we consider a mean-field double layer? The simplest approach is to assume a Helmholtz layer with capacitance $C_{HL}$ and in series to a diffuse layer with capacitance $C_{diffuse}$. In this case, the overall capacitance in a two-electrode configuration can be written as:

$$C_A = \left( \frac{2}{C_{HL}} + \frac{2}{C_{diffuse}} \right)^{-1} = \frac{\varepsilon_0 \cdot \varepsilon_{eff}}{2(r_{Na^+} + L_D)} \quad (5)$$

The radius of the sodium ions is $r_{Na^+} \approx 1 \overset{o}{A}$. The number density of $Na^+$ ions in our glasses is $N_{V,Na^+} \approx 10^{22}\ cm^{-3}$, yielding a Debye length $L_D = \sqrt{\frac{\varepsilon_0 \varepsilon_{eff} k_B T}{N_{V,Na^+} \cdot e^2}} \approx 0.5\ \overset{o}{A}$. When we assume that the effective dielectric constant $\varepsilon_{eff}$ is identical to the electronic polarisability of the sodium ions, we can estimate that $\varepsilon_{eff} \approx 2$. This would result in $C_A \approx 6\ \mu F/cm^2$. When we assume that $\varepsilon_{eff}$ is identical to the macroscopic high-frequency permittivity of the glass, $\varepsilon \approx 10$, we obtain a linear double layer capacitance not higher than $C_A \approx 30\ \mu F/cm^2$. The experimentally determined double layer capacitances of ILs are indeed in this range, but the capacitances of our glasses are considerably higher. One may now argue that the distinction between a Helmholtz and a diffuse layer is somewhat arbitrary when $L_D < r_{Na^+}$. This is certainly correct, but whatever is the exact structure of the double layer, it is hard to believe that its effective thickness is smaller than 1-2 Å and that the effective permittivity in the double layers is higher than 10.

In addition, we considered the possibility that the true interfacial area between our glasses and the Pt electrodes might be higher than the nominal area. In order to get information about this, we studied the surface topography of our glass samples before sputtering the Pt electrodes by means of atomic force microscopy (AFM). However for all glasses, we found that the true surface area is not more than 10% larger than the nominal one.

Consequently, we state that mean-field theories are not sufficient to describe the glass / Pt interfaces. In liquid electrochemistry, capacitance values exceeding $100\ \mu F/cm^2$ have been found in the case of pseudocapacitive charge storage, i.e. when molecules or ions are specifically adsorbed at electrodes [4, 22, 23]. To our knowledge, such processes have not been studied or discussed in any detail for glassy electrolyte materials. In the case of our $Na^+$ ion conducting glasses, the simplest approach is to assume a potential-dependent Langmuir-type adsorption of $Na^+$ ions at the platinum electrodes, which can be described by a Nernst-type equation [4, 23]:

$$\Delta\varphi = \Delta\varphi_0 + \frac{RT}{F} \cdot \ln\left(\frac{1-\vartheta}{\vartheta}\right) \qquad (6)$$

Here, $\Delta\varphi_0$ denotes a standard potential, while $\vartheta$ is the fractional coverage of an electrode by $Na^+$ ions. When $A_{eff}$ is the effective area needed by a single $Na^+$ ion at the electrode surface, then the electrode charge per unit area is given by:

$$\sigma = -\frac{e}{A_{eff}} \cdot \vartheta \qquad (7)$$

This results in the following expression for the differential capacitance:

$$C_A = \left(\frac{d\Delta\varphi}{d\sigma}\right)^{-1} = -\frac{e}{A_{eff}}\left(\frac{d\Delta\varphi}{d\vartheta}\right)^{-1} = \frac{e}{A_{eff}}\left(\frac{RT}{F} \cdot \frac{1}{\vartheta(1-\vartheta)}\right)^{-1} = \frac{e^2}{A_{eff} \cdot k_B T} \cdot \vartheta \cdot (1-\vartheta) \qquad (8)$$

This equation implies that the differential capacitance exhibits a maximum at $\vartheta = 0.5$. In order to estimate the factor $\frac{e^2}{A_{eff} \cdot k_B T}$ we assume that $A_{eff} = (5\ \overset{0}{A})^2$, i.e., due the repulsive interaction between the adsorbed $Na^+$ ions, it is unlikely that the ions approach each other closer than a few $\overset{0}{A}$. At T = 573 K this results in $\frac{e^2}{A_{eff} \cdot k_B T} \approx 1200\ \mu F/cm^2$ and in $C_A^{max} = C_A(\vartheta = 0.5) \approx 300\ \mu F/cm^2$, which is in reasonable agreement with our results shown in Fig. 4. At highly positive and highly negative electrode potentials, we have $\vartheta \to 0$ and $\vartheta \to 1$, respectively, resulting in $C_A \to 0$. Since adsorption phenomena exhibit, in general, a strong temperature dependence, it is not surprising that a change in the sample temperature leads to a thermal relaxation of the measured interfacial capacitance, as shown in Fig. 2.

In our present experiments, we used a configuration with two identical blocking electrodes. Therefore, we have no information about the individual potential drops at both electrodes, and only a semi-quantitative comparison between experiment and theory can be made. In order to learn more about the pseudocapacitive processes at the individual electrodes, we suggest two possibilities: (i) Usage of one blocking and one non-blocking electrode. In this case, the voltage drop at the non-blocking electrode should be negligible. (ii) Usage of one small-area and one large-area electrode. In this case the voltage at the large-area electrode should be negligible. This will be the subject of future experiments.

## 4. Conclusions

We have studied the differential capacitance of different sodium ion conducting glasses between blocking Pt electrodes. We find that the linear interfacial capacitances are considerably higher than expected from mean-field double layer theories and also considerably higher than found for ionic liquids with similar ion concentrations. With increasing bias voltage, the differential interfacial capacitance exhibits a maximum around 1 V and a strong drop at higher voltages. We explain these features in a semi-quantitative fashion by assuming pseudocapacitive charge storage, namely adsorption of $Na^+$ ions at the electrodes. From an application point of view, our results point to the possibility to store a considerable amount of electrical energy at the interface between metals and ion conducting glasses.

**Acknowledgments**
We are grateful to the Alexander von Humboldt Foundation and to German Science Foundation (DFG) for financial support.

**Figure captions:**

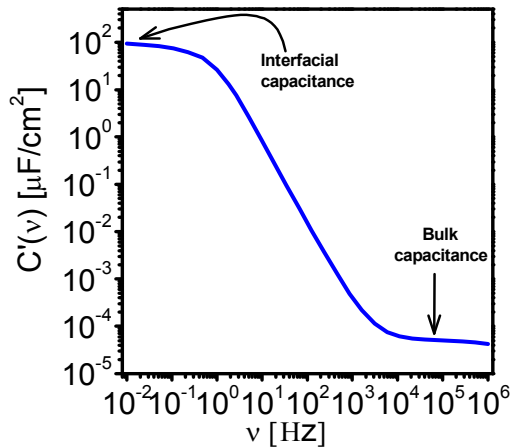

**FIG. 1: Frequency-dependent real part of the capacitance per unit area, $C'(\nu)$, for the 46S4 glass at T = 573 K. The spectrum was taken at zero dc bias voltage using a rms ac voltage of 10 mV.**

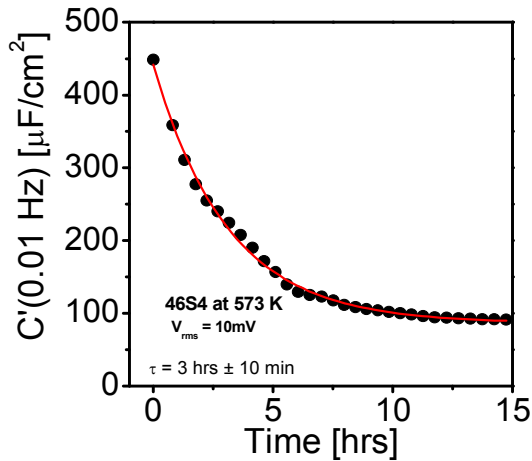

**FIG. 2: Relaxation of the low-frequency capacitance of the 46S4 glass, after heating up a freshly prepared sample to 573K.**

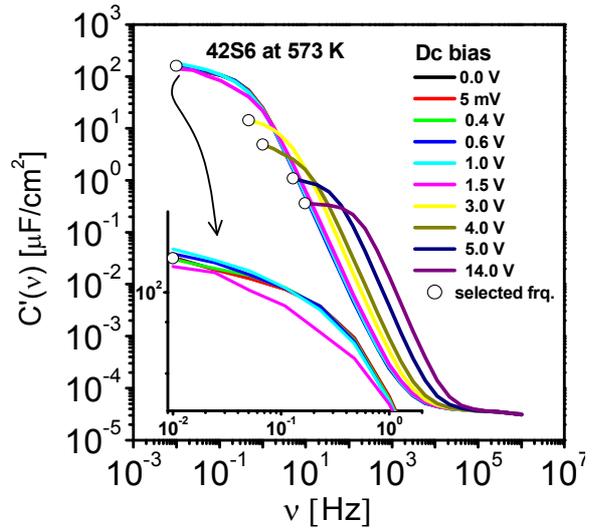

**FIG. 3: Dc bias voltage dependence of the capacitance spectra of the 42S6 glass at T = 573 K. The open circles denote the frequencies selected for plotting the voltage dependence of the differential interfacial capacitance in Fig. 4.**

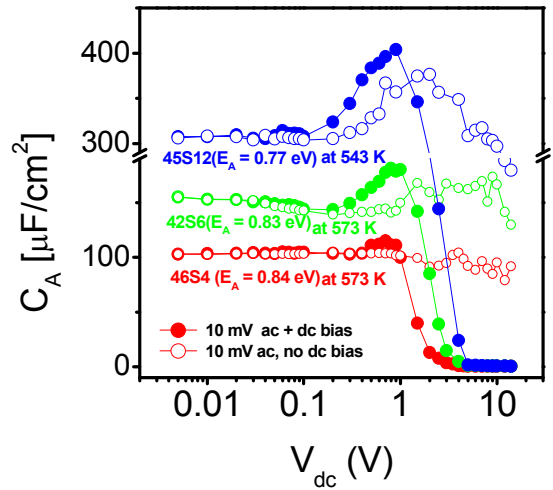

**FIG. 4: Dc bias voltage dependence of the differential interfacial capacitance, $C_A$, for the three glasses 46S4, 42S6, and 45S12. The closed symbols represent the capacitances measured under applied bias voltage $V_{dc}$. The open symbols represent the capacitances measured at zero bias voltage, directly after having performed a measurement at $V_{dc}$.**